\documentclass[12pt]{article}
\usepackage{graphicx,epsfig,amsmath,amsbsy,amssymb,fullpage,citesort}
\newtheorem{THEO}{Theorem}

\begin{document}

\title{
Bayesian Compressive Sensing\\ via Belief Propagation}
\author{\large \sl Dror Baron,$^{\textrm{1}}$ Shriram Sarvotham,$^{\textrm{2}}$  and Richard G.\
Baraniuk\:$^{\textrm{3}}$$\:$\footnote{This work was supported by the
grants NSF CCF-0431150 and CCF-0728867, DARPA/ONR N66001-08-1-2065, ONR N00014-07-1-0936 and
N00014-08-1-1112, AFOSR FA9550-07-1-0301, ARO MURI W311NF-07-1-0185,
and the Texas Instruments Leadership University Program. A preliminary version of this
work appeared in the technical report~\cite{sudoLDPC}.
\protect\\ E-mail: drorb@ee.technion.ac.il, \{shri, richb\}@rice.edu;
Web: dsp.rice.edu/cs}\\[10pt]
\small $^{\textrm{1}}$Department of Electrical Engineering, Technion -- Israel Institute of Technology; Haifa, Israel\\[-0pt]
\small $^{\textrm{2}}$Halliburton; Houston, TX\\[-0pt]
\small $^{\textrm{3}}$Department of Electrical and Computer Engineering, Rice University; Houston, TX\\[-0pt]
}
\maketitle \thispagestyle{empty}

\begin{abstract}
Compressive sensing (CS) is an emerging field based on the revelation that a
small collection of linear projections of a sparse signal contains enough
information for stable, sub-Nyquist signal acquisition. When a statistical characterization
of the signal is available, Bayesian inference can complement conventional CS methods
based on linear programming or greedy algorithms.
We perform approximate Bayesian inference using belief propagation (BP)
decoding, which represents the CS encoding matrix as a graphical model.
Fast computation is obtained by reducing the size of the graphical model
with sparse encoding matrices.
To decode a length-$N$ signal containing $K$ large coefficients,
our CS-BP decoding algorithm uses $O(K\log(N))$ measurements and
$O(N\log^2(N))$ computation. Finally, although we focus on a two-state
mixture Gaussian model, CS-BP is easily adapted to other signal models.
\end{abstract}

\section{Introduction}

Many signal processing applications require the identification and
estimation of a few significant coefficients from a high-dimensional vector.
The wisdom behind this is the ubiquitous compressibility of signals:
in an appropriate basis,
most of the information contained in a signal often resides in just a few large
coefficients. Traditional sensing and processing first acquires the entire
data, only to later throw away most coefficients and retain the few
significant ones~\cite{devore92im}.
Interestingly, the information contained in the few large
coefficients can be captured (encoded) by a small number of random linear
projections~\cite{GR1997}.
The ground-breaking work in {\em compressive sensing}
(CS)~\cite{CandesRUP,DonohoCS,RichBCS2007} has proved for a variety of settings
that the signal can then be decoded in a computationally feasible
manner from these random projections.

\subsection{Compressive sensing}

{\bf Sparsity and random encoding}:\
In a typical compressive sensing (CS) setup, a signal vector
$x\in\mathbb{R}^N$ has the form $x=\Psi\theta$, where
$\Psi\in\mathbb{R}^{N\times N}$ is an orthonormal basis, and
$\theta\in\mathbb{R}^N$ satisfies $\|\theta\|_0=K \ll N$.\footnote{
We use $\|\cdot\|_0$ to denote the $\ell_0$ ``norm" that counts
the number of non-zero elements.}
Owing to the {\em sparsity} of $x$ relative to the basis $\Psi$, there is
no need to sample all $N$ values of $x$. Instead, the CS theory establishes
that $x$ can be decoded from a small number of projections onto an incoherent
set of measurement vectors~\cite{CandesRUP,DonohoCS}. To {\em measure} (encode) $x$,
we compute $M\ll N$ linear projections of $x$ via the matrix-vector
multiplication $y=\Phi x$ where $\Phi\in\mathbb{R}^{M\times N}$ is the
{\em encoding matrix}.

In addition to {\em strictly sparse} signals where $\|\theta\|_0 \leq K$,
other {\em signal models} are possible.
{\em Approximately sparse} signals have $K \ll N$ large
coefficients, while the remaining coefficients are small but not necessarily
zero. {\em Compressible} signals have coefficients that, when sorted,
decay quickly according to a power law.
Similarly, both {\em noiseless} and {\em noisy} signals and measurements may be considered.
We emphasize noiseless measurement of approximately sparse signals
in the paper.

{\bf Decoding via sparsity}:\
Our goal is to decode $x$ given $y$ and $\Phi$.
Although decoding $x$ from $y=\Phi x$ appears to be an ill-posed
inverse problem, the prior knowledge of sparsity in $x$ enables to decode
$x$ from $M \ll N$ measurements. Decoding often relies on an optimization,
which searches for the sparsest coefficients $\theta$ that agree with the
measurements $y$. If $M$ is sufficiently large and $\theta$ is strictly
sparse, then $\theta$ is the solution to the $\ell_0$ minimization:
\[
\widehat{\theta} = \arg\min \|\theta\|_0 ~~~\mbox{s.t. }
y=\Phi \Psi \theta.
\]
Unfortunately, solving this $\ell_0$ optimization is NP-complete~\cite{CandesECLP}.

The revelation that supports the CS theory is that a computationally
tractable optimization problem yields an equivalent solution. We need only solve
for the $\ell_1$-sparsest coefficients that agree with the measurements
$y$~\cite{CandesRUP,DonohoCS}:
\begin{equation}
\widehat{\theta} = \arg\min \|\theta\|_1 ~~~\mbox{s.t. }
y=\Phi \Psi \theta,
\label{eq:L1}
\end{equation}
as long as $\Phi\Psi$ satisfies some technical conditions, which are satisfied
with overwhelming probability when the entries of $\Phi$ are independent and
identically distributed (iid) sub-Gaussian random variables~\cite{CandesRUP}.
This $\ell_1$ optimization problem (\ref{eq:L1}), also known as {\em Basis
Pursuit}~\cite{DonohoBP}, can be solved with linear programming methods.
The $\ell_1$ decoder requires only $M=O(K\log(N/K))$
projections~\cite{DonohoMar2005,DonohoJan2005}. However,
encoding by a dense Gaussian $\Phi$ is slow, and $\ell_1$ decoding requires
cubic computation in general~\cite{Vaidya87}.

\subsection{Fast CS decoding}

While $\ell_1$ decoders figure prominently in the CS literature,
their cubic complexity still renders them impractical
for many applications. For example, current digital cameras acquire images
with $N=10^6$ pixels or more, and fast decoding is critical.
The slowness of $\ell_1$
decoding has motivated a flurry of research into faster algorithms.

One line of research involves iterative greedy algorithms. The {\em Matching Pursuit}
(MP)~\cite{TroppOMP} algorithm, for example, iteratively selects the vectors
from the matrix $\Phi\Psi$ that contain most of the energy of the measurement
vector $y$. MP has been proven to successfully decode the acquired signal
with high probability~\cite{TroppOMP,CDDNOA}. Algorithms inspired by MP include OMP~\cite{TroppOMP},
tree matching pursuit~\cite{MarcoSPARS05}, stagewise OMP~\cite{stomp},
CoSaMP~\cite{Cosamp08}, IHT~\cite{BlumensathDavies2008},
and Subspace Pursuit~\cite{SubspacePursuit08} have been shown to attain similar
guarantees to those of their optimization-based
counterparts~\cite{FPC2007,GPSR2007,BergFriedlander:2008}.

While the CS algorithms discussed above typically use a dense $\Phi$ matrix,
a class of methods has emerged that employ {\em structured} $\Phi$.
For example, subsampling an orthogonal basis that admits a fast implicit algorithm
also leads to fast decoding~\cite{CandesRUP}.
Encoding matrices that are themselves sparse can also be used.
Cormode and Muthukrishnan proposed fast streaming algorithms based on group
testing~\cite{Muthu_old,Muthu}, which considers subsets of signal
coefficients in which we expect at most one ``heavy hitter"
coefficient to lie. Gilbert et~al.~\cite{Gilbert} propose the Chaining
Pursuit algorithm, which works best for extremely sparse signals.

\subsection{Bayesian CS}

CS decoding algorithms rely
on the sparsity of the signal $x$. In some applications,
a statistical characterization of the signal is available, and Bayesian
inference offers the potential for more precise estimation of $x$ or
a reduction in the number of CS measurements. Ji et al.~\cite{BCS2008}
have proposed a Bayesian CS framework where relevance vector machines
are used for signal estimation. For certain types of hierarchical priors,
their method can approximate the posterior density of $x$ and is somewhat
faster than $\ell_1$ decoding.
Seeger and Nickisch~\cite{BCSEx2008} extend these ideas to experimental
design, where the encoding matrix is designed sequentially based on
previous measurements.
Another Bayesian approach by Schniter et al.~\cite{FBMP2009} approximates
conditional expectation by extending the maximal likelihood approach
to a weighted mixture of the most likely models.
There are also many related results on application of Bayesian methods to sparse
inverse problems (c.f.~\cite{Hastie2001} and references therein).

Bayesian approaches have also been used for {\em multiuser decoding} (MUD)
in communications.
In MUD, users modulate their symbols with different spreading
sequences, and the received signals are superpositions of sequences. Because
most users are inactive, MUD algorithms extract information from a sparse
superposition in a manner analogous to CS decoding. Guo and
Wang~\cite{GuoWang2008} perform MUD using sparse spreading sequences and
decode via {\em belief propagation}
(BP)~\cite{Pearl88,Jensen96,Frey98,Yedidia2001,MacKay03,CDLS03};
our paper also uses sparse encoding matrices and BP decoding.
A related algorithm for decoding {\em low density lattice codes}
(LDLC) by Sommer et al.~\cite{LDLC2008} uses BP on a factor
graph whose self and edge potentials are Gaussian mixtures.
Convergence results for the LDLC decoding algorithm have been derived
for Gaussian noise~\cite{LDLC2008}.

\subsection{Contributions}

In this paper, we develop a sparse encoder matrix $\Phi$ and a belief propagation (BP) decoder
to accelerate CS encoding and decoding under the Bayesian framework.
We call our algorithm CS-BP. Although we emphasize a two-state mixture Gaussian
model as a prior for sparse signals, CS-BP is flexible to variations in
the signal and measurement models.

{\bf Encoding by sparse CS matrix}:\
The dense sub-Gaussian CS encoding matrices~\cite{CandesRUP,DonohoCS}
are reminiscent of Shannon's random code constructions. However, although
dense matrices capture the information content of sparse signals,
they may not be amenable to fast encoding and decoding.
{\em Low density parity check} (LDPC) codes~\cite{Gallager62,RSU2001}
offer an important insight: encoding and decoding are fast, because
multiplication by a sparse matrix
is fast; nonetheless, LDPC codes achieve rates close to the Shannon limit.
Indeed, in a previous
paper~\cite{sudo_isit}, we used an LDPC-like sparse $\Phi$ for the special
case of noiseless measurement of strictly
sparse signals; similar matrices were also proposed for CS
by Berinde and Indyk~\cite{BerindeIndyk2008}.
Although LDPC decoding algorithms may not have provable convergence, the recent
extension of LDPC to LDLC codes~\cite{LDLC2008} offers provable convergence,
which may lead to similar future results for CS decoding.

We encode (measure) the signal using sparse
Rademacher ($\{0,1,-1 \}$) LDPC-like $\Phi$ matrices.
Because entries of $\Phi$ are restricted to $\{ 0,1,-1\}$, encoding
only requires sums and differences of small subsets of coefficient
values of $x$. The design of $\Phi$, including characteristics such as
column and row weights, is based on the relevant signal and measurement
models, as well as the accompanying decoding algorithm.

{\bf Decoding by BP}:\
We represent the sparse $\Phi$ as a sparse bipartite graph.
In addition to accelerating the algorithm, the sparse structure
reduces the number of loops in the graph and thus assists the convergence of
a message passing method that solves a Bayesian inference problem.
Our estimate for $x$ explains the measurements
while offering the best match to the
prior. We employ BP in a manner similar to LDPC channel
decoding~\cite{RSU2001,Gallager62,MacKay03}. To decode a length-$N$ signal
containing $K$ large coefficients, our CS-BP decoding algorithm uses $M=O(K\log(N))$
measurements and $O(N\log^2(N))$ computation.
Although CS-BP is not guaranteed to converge, numerical results are
quite favorable.

The remainder of the paper is organized as follows. Section~\ref{sec:sig_model}
defines our signal model, and Section~\ref{subsec:LDPC_Phi} describes our
sparse CS-LDPC encoding matrix.
The CS-BP decoding algorithm is described in Section~\ref{sec:decoding2},
and its performance is demonstrated numerically in Section~\ref{sec:sims}.
Variations and applications are discussed in Section~\ref{sec:variations},
and~Section~\ref{sec:discussion} concludes.

\section{Mixture Gaussian signal model}
\label{sec:sig_model}

We focus on a two-state mixture Gaussian model~\cite{PKH96,CKM97,Crouse98}
as a prior that succinctly captures our prior knowledge about
approximate sparsity of the signal.
Bayesian inference using a two-state mixture model has been studied well
before the advent of CS, for example
by George and McCulloch~\cite{George93} and Geweke~\cite{Geweke96};
the model was proposed for CS in~\cite{sudoLDPC}
and also used by He and Carin~\cite{BCS2008spike}.
More formally, let $X=[X(1),\ldots,X(N)]$ be a random vector in $\mathbb{R}^N$,
and consider the signal $x=[x(1),\ldots,x(N)]$ as an outcome of $X$.
Because our approximately sparse signal consists of a small number of
large coefficients and a large number of small coefficients, we associate
each probability density function (pdf) $f(X(i))$ with a state variable $Q(i)$ that can take
on two values. Large and small magnitudes correspond to zero mean Gaussian
distributions with high and low variances, which are
implied by $Q(i)=1$ and $Q(i)=0$, respectively,
\[
f(X(i) | Q(i)=1)  \sim \mathcal{N}(0,\sigma_1^2) ~~\mbox{and}~~
f(X(i) | Q(i)=0)  \sim \mathcal{N}(0,\sigma_0^2),
\]
with $\sigma_1^2 > \sigma_0^2$. Let $Q=[Q(1),\ldots,Q(N)]$ be
the state random vector associated with the signal; the actual configuration
$q=[q(1),\ldots,q(N)] \in \{0,1 \}^N$ is one of $2^N$ possible outcomes.
We assume that the $Q(i)$'s are iid.\footnote{
The model can be extended to capture dependencies between coefficients,
as suggested by Ji et al.~\cite{BCS2008}. }
To ensure that we have approximately $K$ large coefficients,
we choose the probability mass function (pmf) of the state variable $Q(i)$
to be Bernoulli with $\Pr\left(Q(i)=1\right)=S$ and $\Pr\left(Q(i)=0\right)=1-S$,
where $S=K/N$ is the {\em sparsity rate}.

\begin{figure*}[tt]
\begin{center}
\begin{tabular}{ccccc}
~~{$\Pr(Q=0)$} & & ~~~~{$\Pr(Q=1)$} & & ~~~~{} \\
~~{$f(X| Q=0)$} & & ~~~~{$f(X | Q=1)$} & & ~~~~{$f(X)$} \\
\epsfysize = 25mm
\epsffile{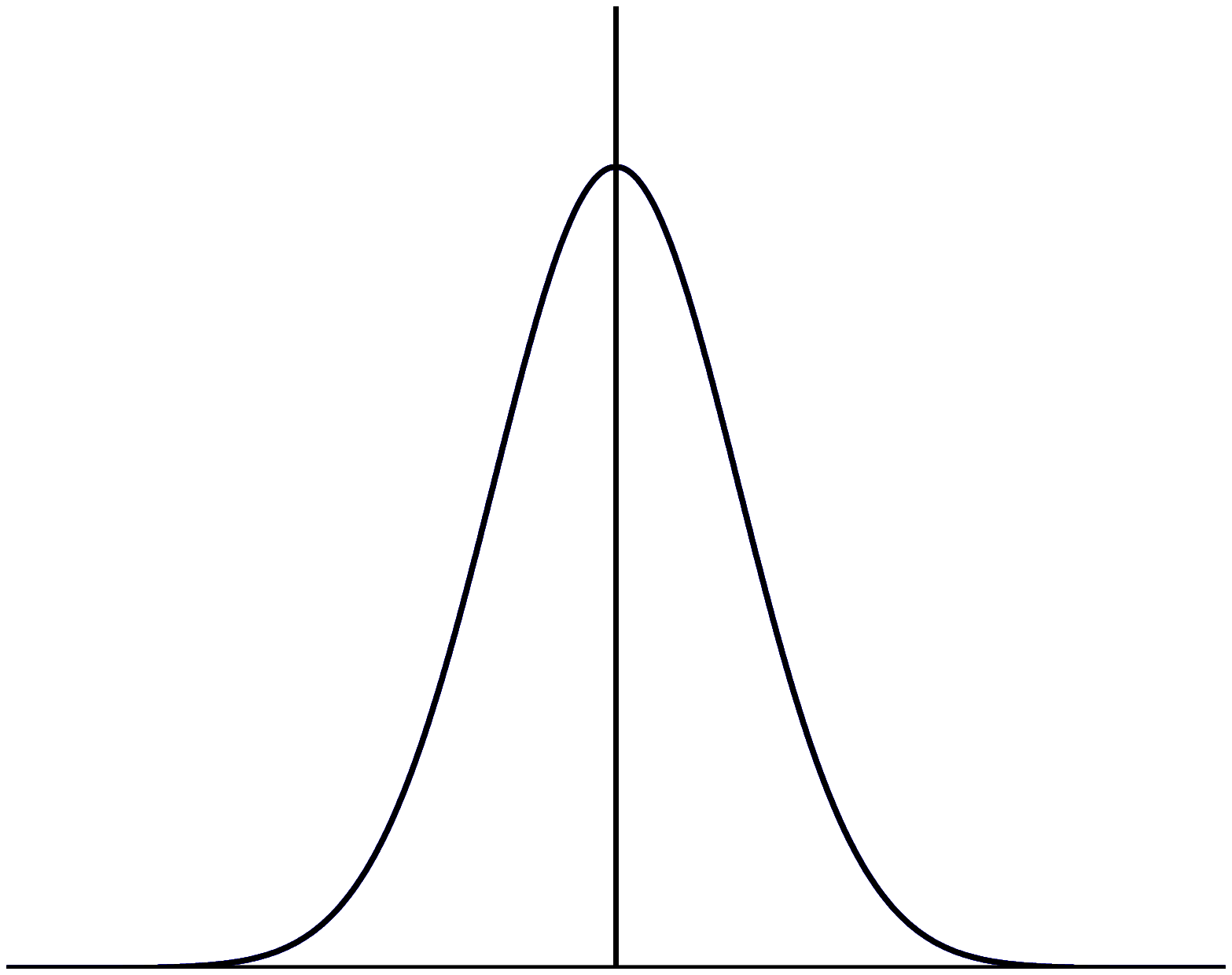} &
\begin{minipage}[t]{5mm}
\vspace{-20mm} \LARGE{ } \vspace{10mm}
\end{minipage} &
\epsfysize = 25mm
\epsffile{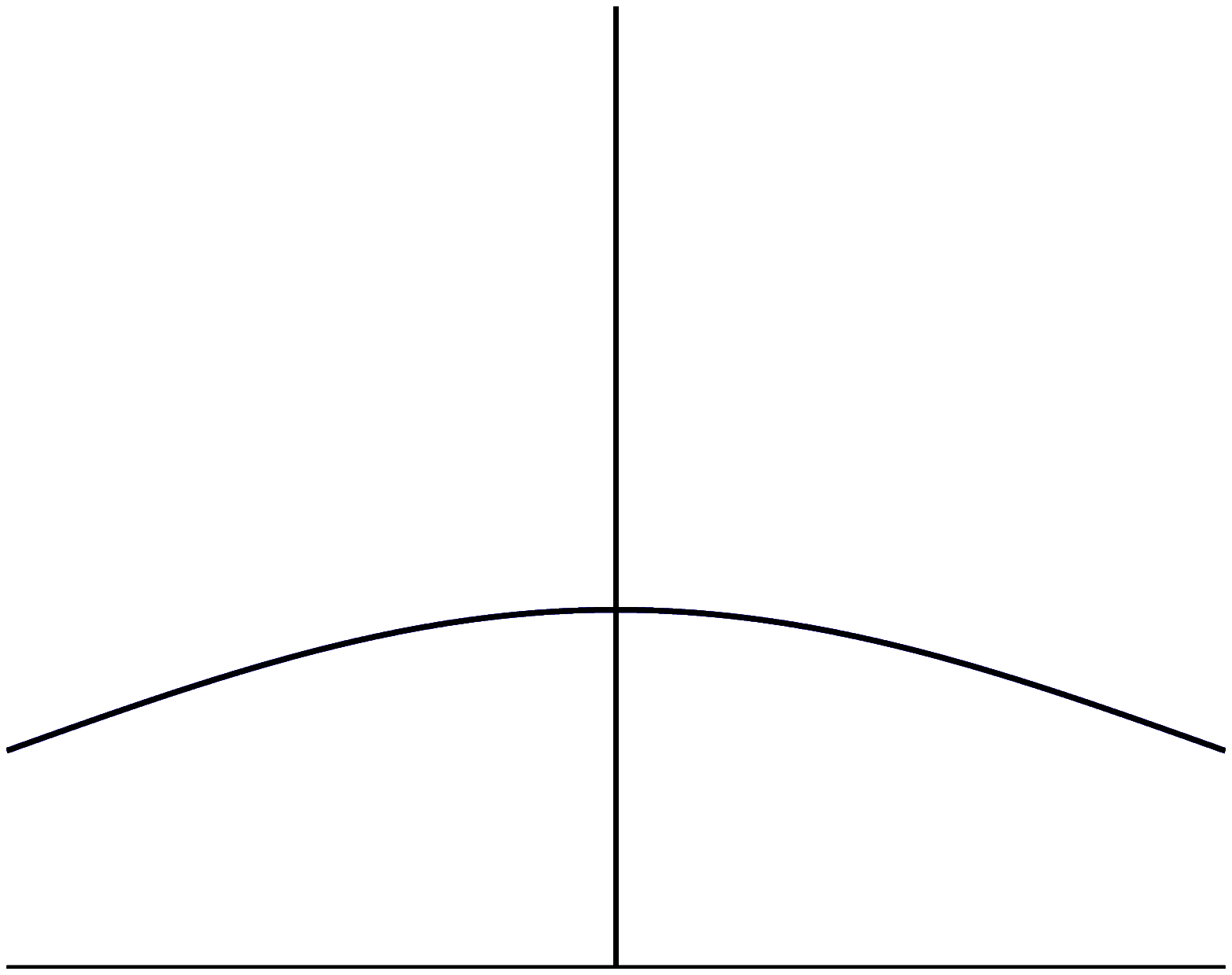} &
\begin{minipage}[t]{5mm}
\vspace{-20mm} \LARGE{$\Rightarrow$} \vspace{10mm}
\end{minipage} &
\epsfysize = 25mm
\epsffile{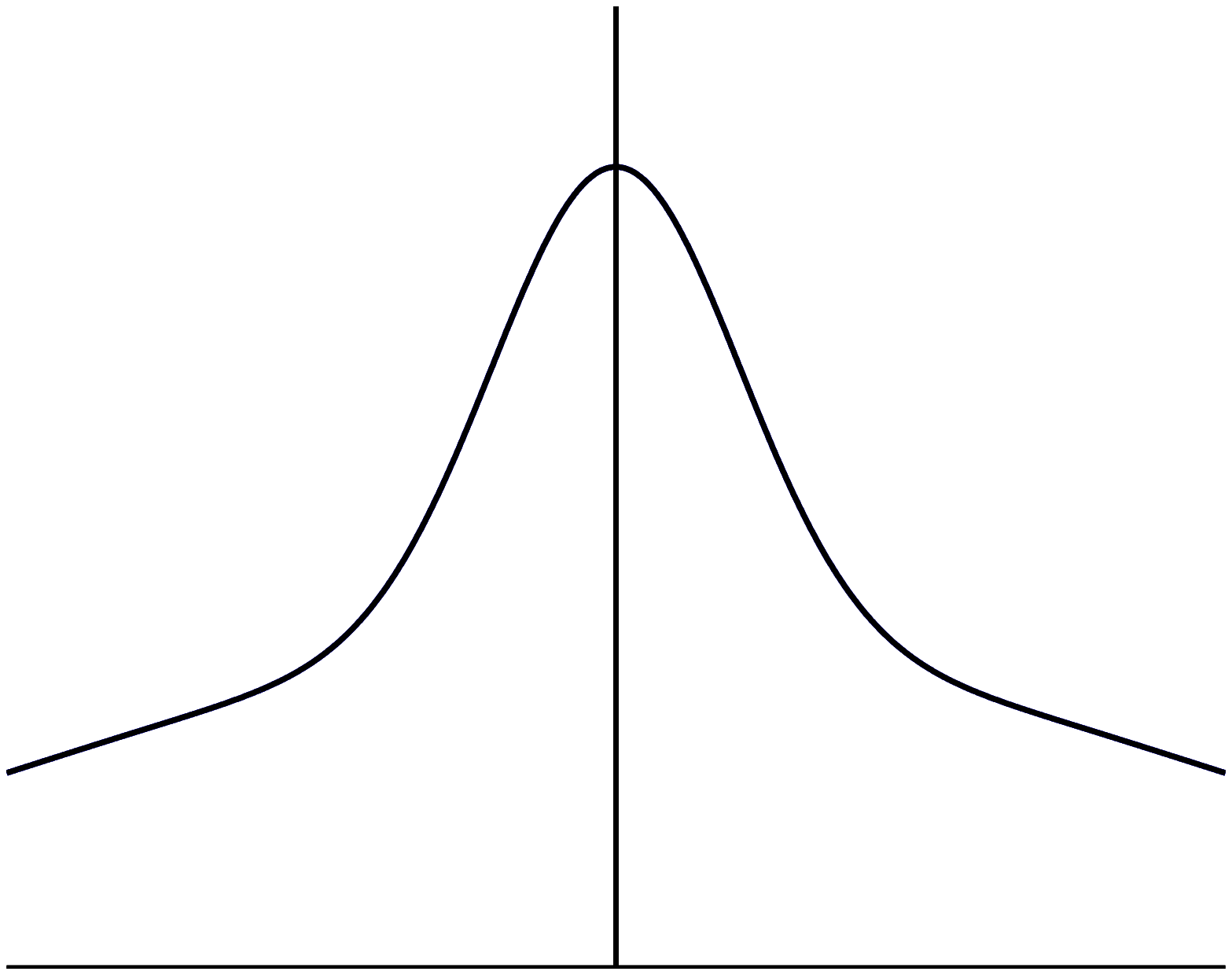} \\
\end{tabular}
\end{center}
\caption{ {\small\sl Mixture Gaussian model for signal coefficients.
The distribution of $X$ conditioned on the two state variables,
$Q=0$ and $Q=1$, is depicted. Also shown is the overall distribution for $X$.
}}
\label{fig:mixmodel}
\end{figure*}

The resulting model for signal coefficients is a two-state mixture Gaussian
distribution, as illustrated in Figure~\ref{fig:mixmodel}.
This mixture model is completely characterized by three
parameters: the sparsity rate $S$ and the variances $\sigma_0^2$ and
$\sigma_1^2$ of the Gaussian pdf's corresponding to each state.

Mixture Gaussian models have been successfully employed in image processing
and inference problems, because they are simple yet effective in modeling
real-world signals~\cite{PKH96,CKM97,Crouse98}. Theoretical connections have
also been made between wavelet coefficient mixture models and the fundamental
parameters of Besov spaces, which have proved invaluable
for characterizing real-world images.  Moreover, arbitrary
densities with a finite number of discontinuities can be approximated
arbitrarily closely by increasing the number of states and allowing
non-zero means~\cite{SA71}. We leave these extensions for future work,
and focus on two-state mixture Gaussian distributions for modeling the
signal coefficients.

\section{Sparse encoding}
\label{subsec:LDPC_Phi}

\begin{figure*}[tt]
\begin{center}
\epsfysize = 60mm
\epsffile{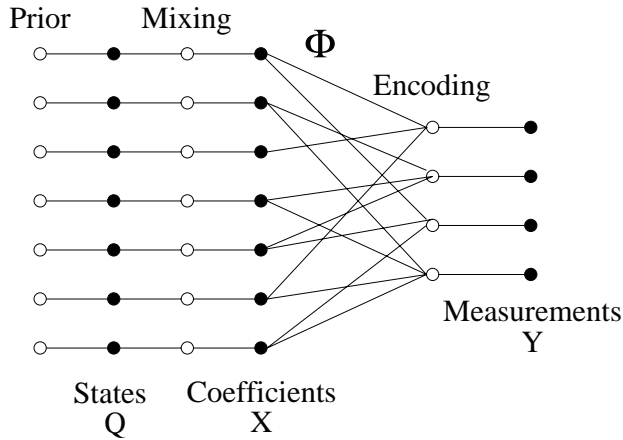}
\end{center}
\vspace*{-5mm}
\caption{{\small\sl Factor graph depicting the relationship between
variable nodes (black) and constraint nodes (white) in CS-BP. }
\label{fig:factorgraph}
}
\end{figure*}

{\bf Sparse CS encoding matrix}:\
We use a sparse $\Phi$ matrix to accelerate both CS encoding and decoding.
Our CS encoding matrices are dominated by zero entries, with a small
number of non-zeros in each row and each column.
We focus on {\em CS-LDPC matrices} whose non-zero
entries are $\{-1,1\}$;\footnote{CS-LDPC matrices are slightly different from LDPC
parity check matrices, which only contain the binary entries $0$
and $1$. We have observed numerically that allowing negative entries offers
improved performance. At the expense of additional computation, further
minor improvement can be attained using sparse matrices with Gaussian non-zero entries.}
each measurement involves only sums and differences of a small subset of
coefficients of $x$.
Although the {\em coherence} between a sparse $\Phi$ and $\Psi$, which is the
maximal inner product between rows of $\Phi$ and $\Psi$, may be higher than
the coherence using a dense $\Phi$ matrix~\cite{Tropp04greed},
as long as $\Phi$ is not too sparse (see Theorem~\ref{th:median_algo} below)
the measurements capture enough information about $x$ to decode the signal.
A CS-LDPC $\Phi$ can be represented as a bipartite graph $G$, which is also
sparse. Each edge of $G$ connects a coefficient node $x(i)$ to an encoding node
$y(j)$ and corresponds to a non-zero entry of $\Phi$ (Figure~\ref{fig:factorgraph}).

In addition to the core structure of $\Phi$,
we can introduce other constraints to tailor the measurement process
to the signal model. The {\em constant row weight} constraint makes sure
that each row of $\Phi$ contains exactly $L$ non-zero entries. The row weight
$L$ can be chosen based on signal properties such as sparsity, possible
measurement noise, and details of the decoding
process. Another option is to use a {\em constant column weight} constraint, which fixes
the number of non-zero entries in each column of $\Phi$ to be a constant $R$.

Although our emphasis is on noiseless measurement of approximately
sparse signals, we briefly discuss noisy measurement of a strictly sparse
signal, and show that a constant row weight $L$ ensures that
the measurements are approximated by two-state mixture Gaussians.
To see this, consider a
strictly sparse $x$ with sparsity rate $S$ and Gaussian variance $\sigma_1^2$.
We now have $y=\Phi x+z$, where $z\sim\cal{N}$$(0,\sigma_Z^2)$ is
{\em additive white Gaussian noise} (AWGN) with
variance $\sigma_Z^2$. In our approximately sparse setting, each row of $\Phi$ picks up
$\approx L(1-S)$ small magnitude coefficients. If $L(1-S)\sigma_0^2\approx \sigma_Z^2$,
then the few large coefficients will be obscured by similar noise artifacts.

Our definition of $\Phi$ relies on the implicit assumption that $x$ is
sparse in the canonical sparsifying basis, i.e., $\Psi=I$. In contrast,
if $x$ is sparse in some other basis $\Psi$, then more complicated encoding
matrices may be necessary. We defer the discussion of these issues to
Section~\ref{sec:variations}, but emphasize that in many practical
situations our methods can be extended to support the sparsifying
basis $\Psi$ in a computationally tractable manner.

{\bf Information content of sparsely encoded measurements}:\
The sparsity of our CS-LDPC matrix may yield measurements $y$ that
contain less information about the signal $x$ than a dense Gaussian $\Phi$.
The following theorem, whose proof appears in the Appendix,
verifies that $y$ retains enough information to decode $x$ well.
As long as $S=K/N =\Omega\left(\left(\frac{\sigma_0}{\sigma_1}\right)^2\right)$,
then $M=O(K\log(N))$ measurements are sufficient.

\begin{THEO} \label{th:median_algo}
Let $x$ be a two-state mixture Gaussian signal
with sparsity rate $S=K/N$ and variances $\sigma_0^2$ and $\sigma_1^2$,
and let $\Phi$ be a CS-LDPC matrix with constant row weight
$L=\eta\frac{\ln(SN^{1+\gamma})}{S}$, where $\eta,\gamma>0$.
If
\begin{equation}
\label{eq:th:median_algo}
M= O\left(
\frac{(1+2\eta^{-1})(1+\gamma)}{\mu^2}\left[2K+(N-K)\left(\frac{\sigma_0}{\sigma_1}\right)^2\right]
\log(N)
\right),
\end{equation}
then $x$ can be decoded to $\widehat{x}$ such that
$\|x-\widehat{x}\|_{\infty} < \mu\sigma_1$ with probability $1-2N^{-\gamma}$.
\end{THEO}

The proof of Theorem~\ref{th:median_algo} relies on a result by
Wang et al.~\cite[Theorem 1]{WGR2007}. Their proof partitions $\Phi$ into
$M_2$ sub-matrices of $M_1$ rows each, and estimates each $\widehat{x_i}$
as a median of inner products with sub-matrices. The $\ell_{\infty}$
performance guarantee relies on the union bound; a less stringent
guarantee yields a reduction in $M_2$. Moreover, $L$ can be reduced if
we increase the number of measurements accordingly. Based on numerical
results, we propose the following modified values as rules of thumb,
\begin{equation}
\label{eq:thumb_LM}
L \approx S^{-1}=N/K, \quad
M=O(K\log(N)), \quad \mbox{and} \quad
R=LM/N=O(\log(N)).
\end{equation}
Noting that each measurement requires $O(L)$ additions
and subtractions, and using our rules of thumb for $L$ and $M$
(\ref{eq:thumb_LM}), the computation required for encoding is $O(LM)=O(N\log(N))$,
which is significantly lower than the $O(MN)=O(KN\log(N))$ required for dense Gaussian $\Phi$.

\section{CS-BP decoding of approximately sparse signals}
\label{sec:decoding2}

Decoding approximately sparse random signals can be treated as a Bayesian inference problem.
We observe the measurements $y=\Phi x$, where $x$ is a mixture Gaussian signal.
Our goal is to estimate $x$ given $\Phi$ and $y$. Because the set of equations
$y=\Phi x$ is under-determined, there are infinitely many
solutions. All solutions lie along a hyperplane of dimension
$N-M$. We locate the solution within
this hyperplane that best matches our prior signal model. Consider
the {\em minimum mean square error} (MMSE) and {\em maximum a posteriori} (MAP)
estimates,
\begin{eqnarray*}
\widehat{x}_{\rm MMSE} &=&  \mbox{arg } \min_{{x}'} E\|X-{x}'\|_2^2 \quad \mbox{s.t.} \quad y=\Phi {x}', \\
\widehat{x}_{\rm MAP} &=& \mbox{arg } \max_{{x}'} f(X=x') \quad \mbox{s.t.} \quad y=\Phi {x}',
\end{eqnarray*}
where the expectation is taken over the prior distribution for $X$. The
MMSE estimate can be expressed as the conditional mean,
$\widehat{x}_{\rm MMSE}=E\left[ X | Y=y  \right]$, where
$Y\in\mathbb{R}^M$ is the random vector that corresponds to the
measurements. Although the precise computation of $\widehat{x}_{\rm MMSE}$
may require the evaluation of $2^N$ terms, a close approximation to the
MMSE estimate can be obtained using the (usually small) set of state
configuration vectors $q$ with dominant posterior probability~\cite{FBMP2009}.
Indeed, exact inference in graphical models is
NP-hard~\cite{Cooper90}, because of loops in the graph induced by $\Phi$.
However, the sparse structure of $\Phi$ reduces the number of loops and enables
us to use low-complexity message-passing methods to estimate $x$ approximately.

\subsection{Decoding algorithm}
\label{sec:subbp}

We now employ belief propagation (BP), an efficient method for solving
inference problems by iteratively passing messages over graphical
models~\cite{Pearl88,Jensen96,Frey98,Yedidia2001,MacKay03,CDLS03}.
Although BP has not been proved to converge, for graphs with few loops
it often offers a good approximation to the solution to the MAP inference problem.
BP relies on {\em factor graphs}, which enable fast computation
of global multivariate functions by exploiting the way in which the global
function factors into a product of simpler local functions, each of which
depends on a subset of variables~\cite{KFL01}.

{\bf Factor graph for CS-BP}:\
The factor graph shown in Figure~\ref{fig:factorgraph} captures the
relationship between the states $q$, the signal coefficients $x$, and the
observed CS measurements $y$. The graph is bipartite and contains
two types of vertices; all edges connect {\em variable} nodes (black)
and {\em constraint} nodes (white). There are three types
of variable nodes corresponding to {\em state} variables $Q(i)$,
{\em coefficient} variables $X(i)$, and {\em measurement} variables
$Y(j)$. The factor graph also has three types of constraint nodes,
which encapsulate the dependencies that their neighbors in the graph
(variable nodes) are subjected to.
First, {\em prior constraint nodes} impose the Bernoulli prior on
state variables.
Second, {\em mixing} constraint nodes impose the conditional distribution
on coefficient variables given the state variables.
Third, {\em encoding} constraint nodes impose the encoding matrix structure
on measurement variables.

{\bf Message passing}:\
CS-BP approximates the marginal distributions
of all coefficient and state variables
in the factor graph, conditioned on the observed measurements $Y$, by passing
messages between variable nodes and constraint nodes. Each message encodes the marginal distributions of a variable
associated with one of the edges.
Given the distributions $\Pr(Q(i) | Y=y)$ and $f(X(i) | Y=y)$,
one can extract MAP and MMSE estimates for each coefficient.

Denote the message sent from a variable node $v$ to one of its neighbors in the bipartite graph,
a constraint node $c$, by $\mu_{v \longrightarrow c}(v)$; a message from
$c$ to $v$ is denoted by $\mu_{c \longrightarrow v}(v)$.
The message $\mu_{v \longrightarrow c}(v)$ is updated by taking the product of
all messages received
by $v$ on all other edges. The message $\mu_{c \longrightarrow v}(v)$
is computed in a similar manner, but the constraint associated
with $c$ is applied to the product and the result is marginalized.
More formally,
\begin{equation}
\mu_{v \longrightarrow c}(v) = \prod_{u \in n(v)\setminus\{c\} } \mu_{u \longrightarrow v}(v),
\label{eq:var_to_constraint}
\end{equation}
\begin{equation}
\mu_{c \longrightarrow v}(v) = \sum_{ \sim\{v\} } \left( \mbox{con}(n(c))
\prod_{w \in n(c)\setminus\{v\} } \mu_{w \longrightarrow c}(w) \right),
\label{eq:constraint_to_var}
\end{equation}
where $n(v)$ and $n(c)$ are sets of neighbors of $v$ and $c$, respectively,
$\mbox{con}(n(c))$ is the constraint on the set of variable nodes $n(c)$,
and $\sim\{v\}$ is the set of neighbors of $c$ excluding $v$.
We interpret these 2 types of message processing as {\em multiplication}
of beliefs at variable nodes (\ref{eq:var_to_constraint}) and
{\em convolution} at constraint nodes (\ref{eq:constraint_to_var}).
Finally, the marginal distribution $f(v)$ for a given variable node is obtained from the
product of all the most recent incoming messages along the edges connecting
to that node,
\begin{equation}
\label{eq:BP_marginal}
f(v) = \prod_{u \in n(v) } \mu_{u \longrightarrow v}(v).
\end{equation}
Based on the marginal distribution, various statistical characterizations
can be computed, including MMSE, MAP, error bars, and so on.

We also need a method to encode beliefs.
One method is to sample the relevant pdf's uniformly and then use the samples as
messages. Another encoding method is to approximate the pdf by a mixture
Gaussian with a given number of components, where mixture parameters are
used as messages. These two methods offer different trade-offs between
modeling flexibility and computational requirements;
details appear in Sections~\ref{subsec:msg_samples}
and \ref{subsec:msg_mixtures}.  We leave alternative methods such as
particle filters and importance sampling for future research.

{\bf Protecting against loopy graphs and message quantization errors}:\
BP converges to the exact conditional distribution in the ideal situation
where the following conditions are met:
({\em i}) the factor graph is cycle-free; and
({\em ii}) messages are processed and propagated without errors.
In CS-BP decoding, both conditions
are violated. First, the factor graph is loopy --- it contains cycles.
Second, message encoding methods introduce errors. These non-idealities may
lead CS-BP to converge to imprecise conditional distributions, or more
critically, lead CS-BP to diverge~\cite{SIFW02,FM98,IFW05}.
To some extent these problems can be reduced by ({\em i}) using CS-LDPC matrices, which
have a relatively modest number of loops; and ({\em ii}) carefully designing
our message encoding methods (Sections~\ref{subsec:msg_samples}
and \ref{subsec:msg_mixtures}).
We stabilize CS-BP against these non-idealities using message damped belief
propagation (MDBP)~\cite{Pretti05}, where messages are
weighted averages between old and new estimates.
Despite the damping, CS-BP is not guaranteed to converge, and yet the
numerical results of Section~\ref{sec:sims} demonstrate that its
performance is quite promising.
We conclude with a prototype algorithm; Matlab code is available at
$\mbox{http://dsp.rice.edu/CSBP}$.

\bigskip\centerline{\bf CS-BP Decoding Algorithm}
\begin{enumerate}
\item
{\bf Initialization}:\
Initialize the iteration counter $i=1$.
Set up data structures for factor graph messages
$\mu_{v \longrightarrow c}(v)$ and
$\mu_{c \longrightarrow v}(v)$.
Initialize messages $\mu_{v \longrightarrow c}(v)$ from variable to
constraint nodes with the signal prior.

\item
\label{algo:backward}
{\bf Convolution}:\
For each measurement $c=1,\ldots,M$, which corresponds
to constraint node $c$, compute
$\mu_{c \longrightarrow v}(v)$ via convolution (\ref{eq:constraint_to_var})
for all neighboring variable nodes $n(c)$.
If measurement noise is present, then convolve further with a noise prior.
Apply damping methods such as MDBP~\cite{Pretti05} by weighting
the new estimates from iteration~$i$ with estimates from previous iterations.

\item
{\bf Multiplication}:\
For each coefficient $v=1,\ldots,N$, which corresponds
to a variable node $v$, compute $\mu_{v \longrightarrow c}(v)$
via multiplication (\ref{eq:var_to_constraint})
for all neighboring constraint nodes $n(v)$.
Apply damping methods as needed.
If the iteration counter has yet to reach its maximal value, then
go to Step~\ref{algo:backward}.

\item
{\bf Output}:\
For each coefficient $v=1,\ldots,N$, compute MMSE or
MAP estimates (or alternative statistical characterizations)
based on the marginal distribution $f(v)$
(\ref{eq:BP_marginal}). Output the requisite statistics.
\end{enumerate}

\subsection{Samples of the pdf as messages}
\label{subsec:msg_samples}

Having described main aspects of the CS-BP decoding
algorithm, we now focus on the two
message encoding methods, starting with samples.
In this method, we sample the pdf and send the samples
as messages. Multiplication of pdf's (\ref{eq:var_to_constraint}) corresponds
to point-wise multiplication of messages; convolution
(\ref{eq:constraint_to_var}) is
computed efficiently in the frequency domain.\footnote{Fast convolution
via FFT has been used in LDPC decoding over $GF(2^q)$ using
BP~\cite{MacKay03}.}

The main advantage of using samples is flexibility to different
prior distributions for the coefficients; for example, mixture Gaussian
priors are easily supported. Additionally, both
multiplication and convolution are computed efficiently.
However, sampling has large memory requirements and introduces quantization
errors that reduce precision and hamper the convergence of CS-BP~\cite{SIFW02}.
Sampling also requires finer sampling for precise decoding;
we propose to sample the pdf's with a spacing
less than $\sigma_0$.

We analyze the computational requirements of this method.
Let each message be a vector of $p$ samples. Each iteration
performs multiplication at coefficient nodes (\ref{eq:var_to_constraint})
and convolution at constraint nodes (\ref{eq:constraint_to_var}).
Outgoing messages are modified,
\begin{equation}
\label{eq:modified}
\mu_{v \longrightarrow c}(v) =
\frac{\prod_{u \in n(v) } \mu_{u \longrightarrow v}(v) } {\mu_{c \longrightarrow v}(v)}
\quad \mbox{and} \quad
\mu_{c \longrightarrow v}(v) =
\sum_{ \sim\{v\} } \left( \mbox{con}(n(c)) \frac{\prod_{w \in n(c) }
\mu_{w \longrightarrow c}(w)}{\mu_{v \longrightarrow c}(v)} \right),
\end{equation}
where the denominators are non-zero, because
mixture Gaussian pdf's are strictly positive. The modifications
(\ref{eq:modified}) reduce computation, because the numerators are computed
once and then reused for all messages leaving the node being processed.

Assuming that the column weight $R$ is fixed (Section~\ref{subsec:LDPC_Phi}),
the computation required for message processing at a variable node is $O(Rp)$
per iteration, because we multiply $R+1$ vectors of length $p$. With $O(N)$
variable nodes, each iteration requires $O(NRp)$ computation. For constraint
nodes, we perform convolution in the frequency domain, and so the computational
cost per node is $O(Lp \log(p))$. With $O(M)$ constraint nodes, each iteration is
$O(LMp \log(p))$. Accounting for both variable and constraint
nodes, each iteration is $O(NRp + LMp \log(p))=O(p\log(p)N\log(N))$,
where we employ our rules of thumb for $L$, $M$, and $R$ (\ref{eq:thumb_LM}).
To complete the computational analysis, we note first that we use $O(\log(N))$
CS-BP iterations, which is proportional to the diameter of the graph~\cite{MacKay99}.
Second, sampling the pdf's with a spacing less than
$\sigma_0$, we choose $p=O(\sigma_1 / \sigma_0)$ to support
a maximal amplitude on the order of $\sigma_1$. Therefore, our overall computation is
$O\left(\frac{\sigma_1}{\sigma_0} \log\left(\frac{\sigma_1}{\sigma_0}\right)N\log^2(N)\right)$,
which scales as $O(N\log^2(N))$ when $\sigma_0$ and $\sigma_1$ are constant.

\subsection{Mixture Gaussian parameters as messages}
\label{subsec:msg_mixtures}

In this method, we approximate the
pdf by a mixture Gaussian with a maximum number of components,
and then send the mixture parameters as messages.
For both multiplication (\ref{eq:var_to_constraint}) and convolution
(\ref{eq:constraint_to_var}), the resulting number of components in
the mixture is multiplicative in the number of constituent components.
To keep the message representation tractable, we perform model
reduction using the {\em Iterative Pairwise Replacement Algorithm}
(IPRA)~\cite{SS01}, where a sequence of mixture models is computed
iteratively.

The advantage of using mixture Gaussians to encode pdf's
is that the messages are short and hence consume
little memory. This method works well for mixture Gaussian priors,
but could be difficult to adapt to other priors. Model order reduction
algorithms such as IPRA can be computationally expensive~\cite{SS01},
and introduce errors in the messages, which impair the quality of the solution
as well as the convergence of CS-BP~\cite{SIFW02}.

Again, we analyze the computational requirements.
Because it is impossible to undo the multiplication in
(\ref{eq:var_to_constraint}) and (\ref{eq:constraint_to_var}),
we cannot use the modified form~(\ref{eq:modified}).
Let $m$ be the maximum model order.
Model order reduction using IPRA~\cite{SS01} requires
$O(m^2  R^2)$ computation per coefficient node per iteration.
With $O(N)$ coefficient nodes, each iteration is $O(m^2  R^2N)$.
Similarly, with $O(M)$ constraint nodes, each iteration is
$O(m^2 L^2 M)$. Accounting for
$O(\log(N))$ CS-BP iterations, overall computation is
$O(m^2[L^2M+R^2N]\log(N))=O\left(m^2\frac{N}{S}\log^2(N)\right)$.

\subsection{Properties of CS-BP decoding}

\begin{table}
\caption[Comparison of CS-BP methods]
{\small\sl Computational and storage requirements of CS-BP decoding}
\centering
\begin{tabular}{|l|l|l|l|}\hline
Messages  &  		Parameter  &  Computation & Storage \\
\hline
Samples of pdf &	$p=O(\sigma_1/\sigma_0)$ samples    &
$O\left(\frac{\sigma_1}{\sigma_0} \log\left(\frac{\sigma_1}{\sigma_0}\right)N\log^2(N)\right)$ &
$O(pN\log(N))$ \\
Mixture Gaussians &	$m$ components & $O\left(m^2\frac{N}{S}\log^2(N)\right)$ &
$O(mN\log(N))$ \\
\hline
\end{tabular}
\label{tbl:comparison}
\end{table}

We briefly describe several properties of CS-BP decoding.
The {\em computational characteristics} of the two methods for encoding
beliefs about conditional distributions were evaluated in
Sections~\ref{subsec:msg_samples} and \ref{subsec:msg_mixtures}.
The {\em storage requirements} are mainly for message representation of
the $LM=O(N\log(N))$ edges. For encoding with pdf samples,
the message length is $p$, and so the storage
requirement is $O(pN\log(N))$. For encoding with mixture Gaussian
parameters, the message length is $m$, and so the storage requirement is
$O(mN\log(N))$. Computational and storage requirements are summarized
in Table~\ref{tbl:comparison}.

Several additional properties are now featured.
First, we have {\em progressive decoding}; more measurements
will improve the precision of the estimated posterior probabilities.
Second, if we are only interested in an estimate of the state configuration vector $q$
but not in the coefficient values, then less information must be
extracted from the measurements. Consequently, the number of
measurements can be reduced.
Third, we have {\em robustness to noise}, because noisy measurements
can be incorporated into our model by convolving
the noiseless version of the estimated pdf (\ref{eq:constraint_to_var})
at each encoding node with the pdf of the noise.

\section{Numerical results}
\label{sec:sims}

\begin{figure*}[tt]
\begin{center}
\begin{tabular}{c}
\epsfysize = 60mm
\epsffile{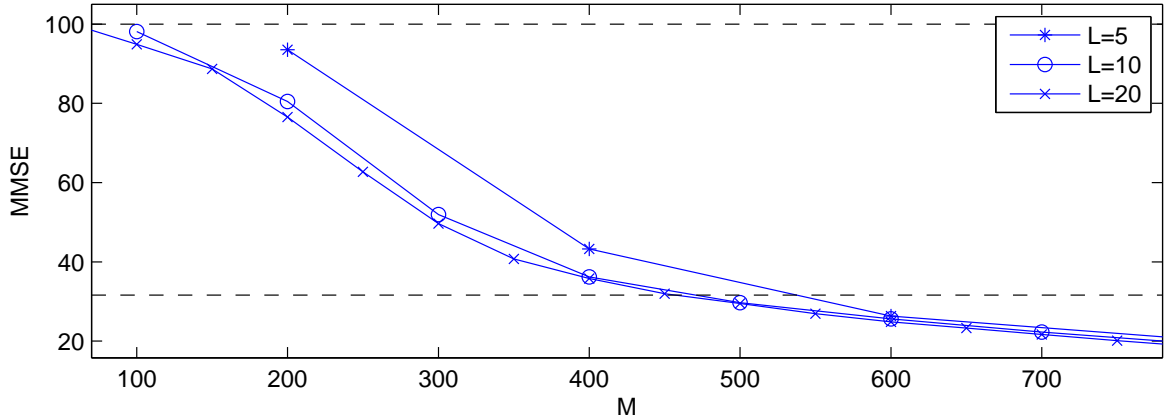}
\end{tabular}
\end{center}
\caption{{\small\sl MMSE as a function of the number of
measurements $M$ using different matrix row weights $L$. The dashed lines show
the $\ell_2$ norms of $x$ (top) and the small coefficients (bottom).
($N=1000$, $S=0.1$, $\sigma_1=10$, $\sigma_0=1$, and noiseless measurements.)}
\label{fig:sample_simresult}
}
\end{figure*}

\begin{figure*}[tt]
\begin{center}
\begin{tabular}{c}
\epsfysize = 80mm
\epsffile{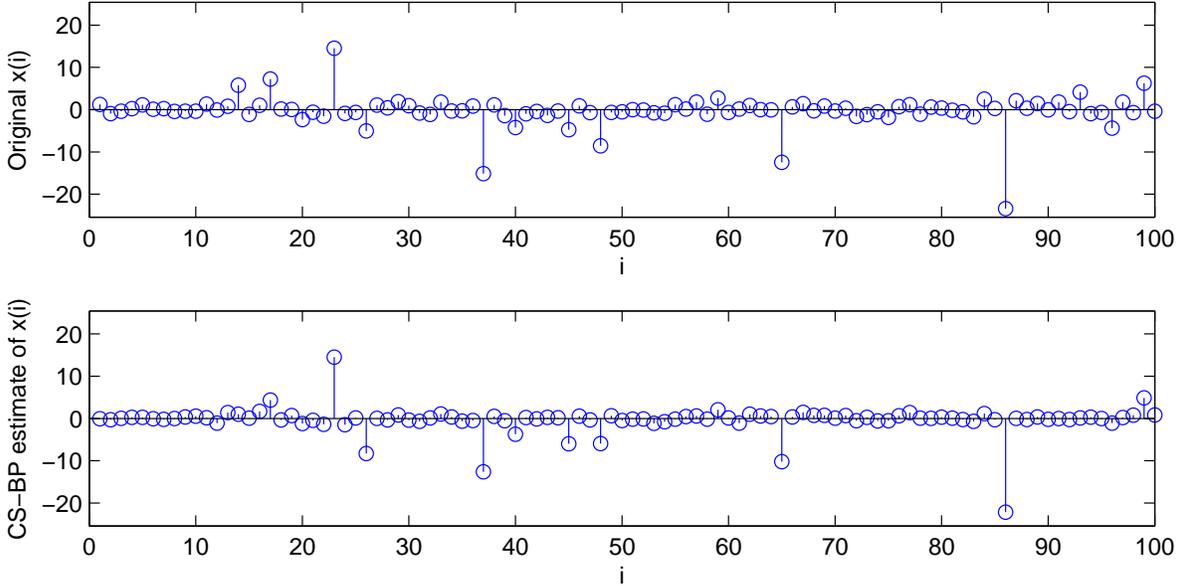}
\end{tabular}
\end{center}
\caption{{\small\sl Original signal $x$ and version decoded by CS-BP.
($N=1000$, $S=0.1$, $L=20$, $M=400$, $\sigma_1=10$, $\sigma_0=1$, and noiseless measurements.)}
\label{fig:sim6_example}
}
\end{figure*}

\begin{figure*}[tt]
\begin{center}
\begin{tabular}{c}
\epsfysize = 60mm
\epsffile{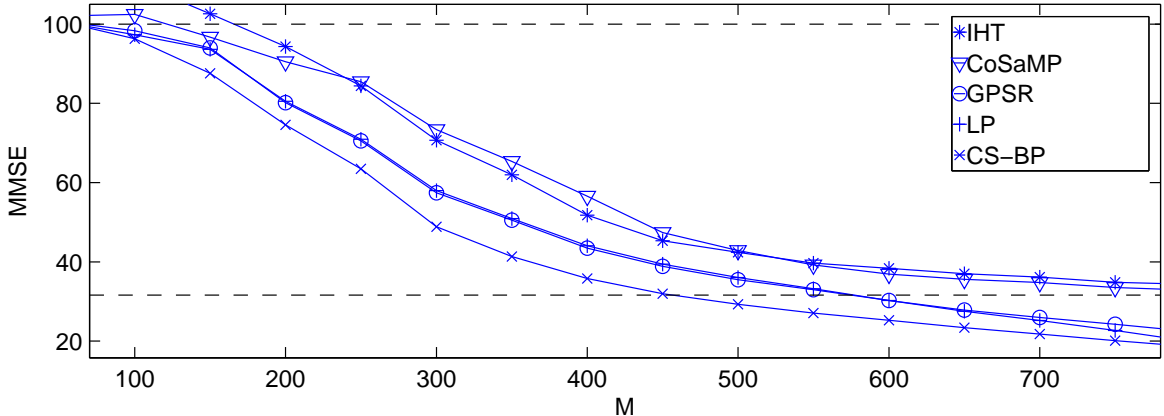}
\end{tabular}
\end{center}
\caption{{\small\sl MMSE as a function of the number of
measurements $M$ using CS-BP, linear programming (LP), GPSR,
CoSaMP, and IHT.
The dashed lines show the $\ell_2$ norms of $x$ (top) and the small coefficients (bottom).
($N=1000$, $S=0.1$, $L=20$, $\sigma_1=10$, $\sigma_0=1$, and noiseless measurements.)}
\label{fig:mse_compare}
}
\end{figure*}

\begin{figure*}[tt]
\begin{center}
\begin{tabular}{c}
\epsfysize = 60mm
\epsffile{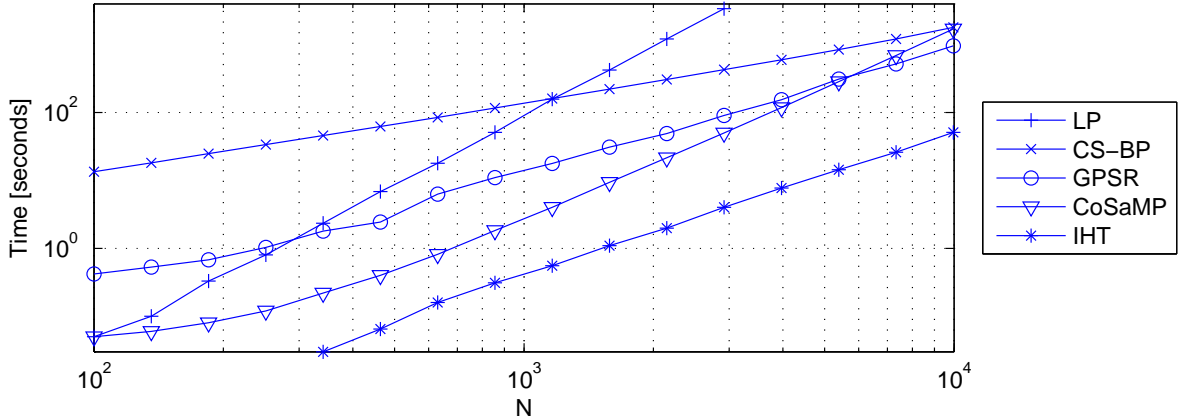}
\end{tabular}
\end{center}
\caption{{\small\sl Run-time in seconds as a function of the
signal length $N$ using CS-BP, linear programming (LP) $\ell_1$ decoding,
GPSR, CoSaMP, and IHT.
($S=0.1$, $L=20$, $M=0.4N$, $\sigma_1=10$, $\sigma_0=1$, and noiseless measurements.)}
\label{fig:timing}
}
\end{figure*}

\begin{figure*}[tt]
\begin{center}
\begin{tabular}{c}
\epsfysize = 60mm
\epsffile{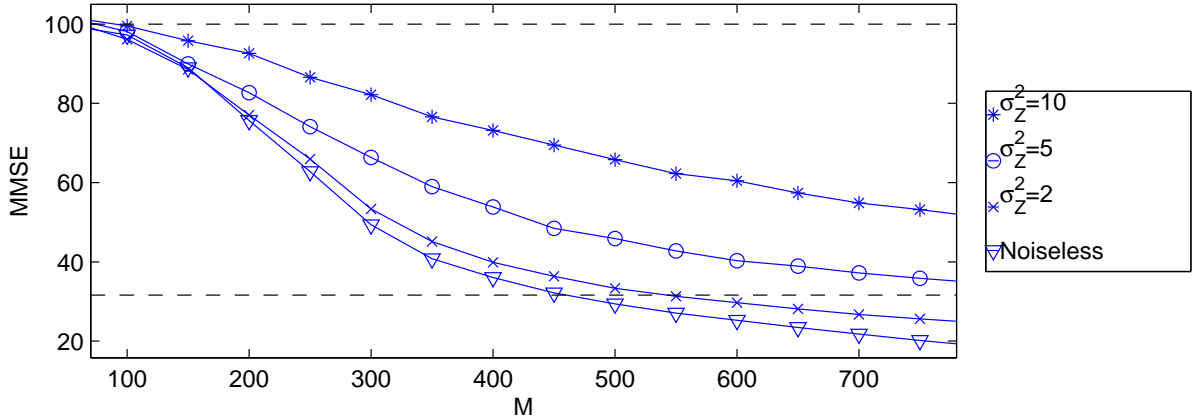}
\end{tabular}
\end{center}
\caption{{\small\sl MMSE as a function of $M$ using different noise levels $\sigma_Z^2$.
The dashed lines show the $\ell_2$ norms of $x$ (top) and the small coefficients (bottom).
($N=1000$, $S=0.1$, $L=20$, $\sigma_1=10$, and $\sigma_0=1$.)}
\label{fig:noisy}
}
\end{figure*}

\begin{figure*}[tt]
\begin{center}
\begin{tabular}{c}
\epsfysize = 110mm
\epsffile{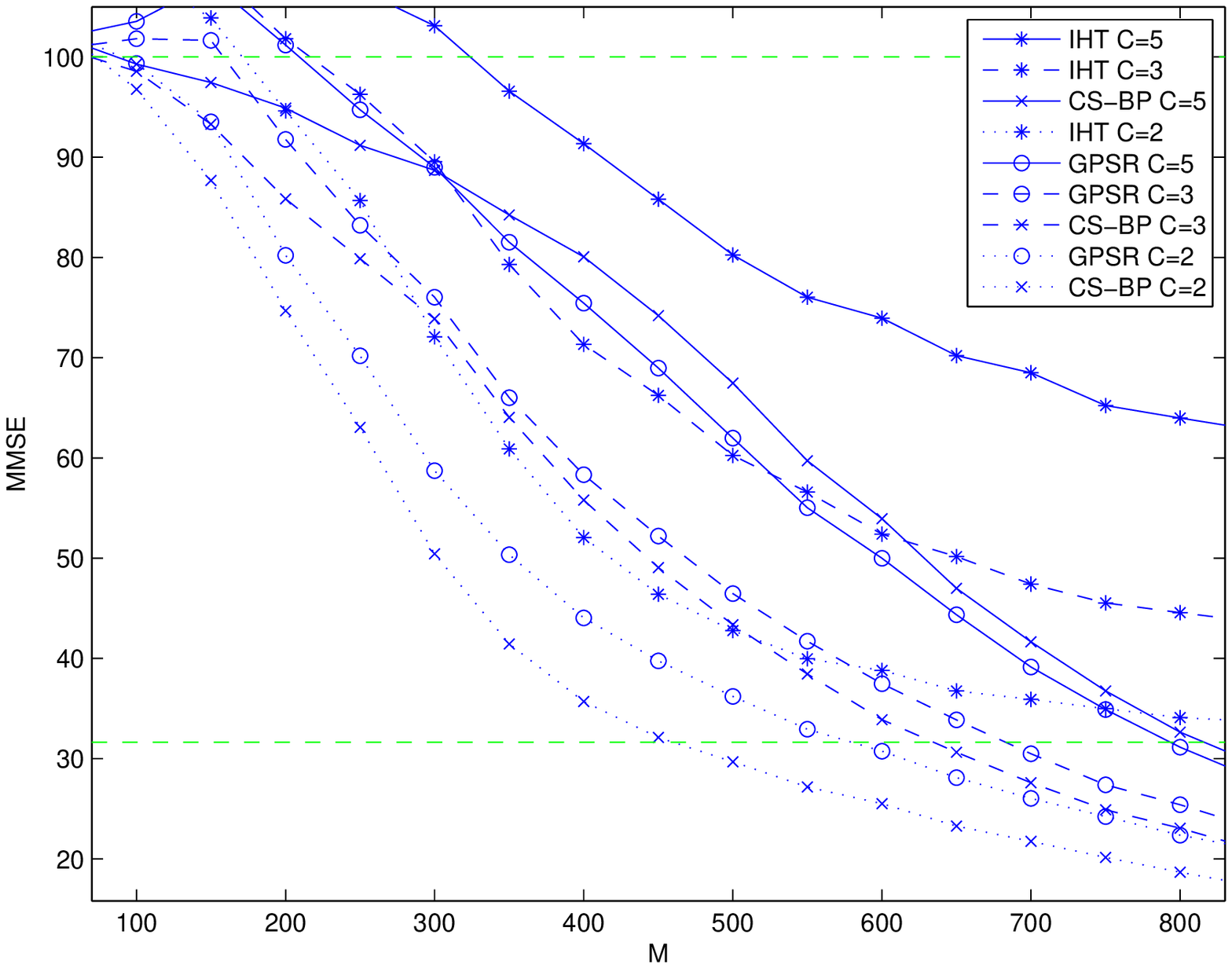}
\end{tabular}
\end{center}
\caption{{\small\sl MMSE as a function of the number of
measurements $M$ and the number of components $C$ in the mixture Gaussian signal model.
Plots for CS-BP (x), GPSR (circle), and IHT (asterisk) appear for
$C=2$ (dotted), $C=3$ (dashed), and $C=5$ (solid).
The horizontal dashed lines show the $\ell_2$ norms of $x$ (top) and the small coefficients (bottom).
($N=1000$, $S=0.1$, $L=20$, $\sigma_1=10$, $\sigma_0=1$, and noiseless measurements.)}
\label{fig:robustness}
}
\end{figure*}

To demonstrate the efficacy of CS-BP, we simulated several different
settings. In our first setting, we considered decoding
problems where $N=1000$, $S=0.1$, $\sigma_1=10$, $\sigma_0=1$, and the
measurements are noiseless. We used samples of the pdf as messages,
where each message consisted of $p=525=3 \cdot 5^2 \cdot 7$ samples;
this choice of $p$ provided fast FFT computation.
Figure~\ref{fig:sample_simresult} plots the MMSE decoding error
as a function of $M$ for a variety of row weights $L$.
The figure emphasizes with dashed lines the average $\ell_2$ norm of $x$
(top) and of the small coefficients (bottom); increasing $M$ reduces the decoding
error, until it reaches the energy level of the small coefficients.
A small row weight, e.g., $L=5$, may miss some of the large coefficients
and is thus bad for decoding; as we increase $L$,
fewer measurements are needed to obtain the same precision. However, there is
an optimal $L_{\rm opt}\approx 2/S=20$ beyond which any performance gains are marginal.
Furthermore, values of $L > L_{\rm opt}$ give rise to divergence in CS-BP,
even with damping.
An example of the output of the CS-BP decoder and how it compares to
the signal $x$ appears in Figure~\ref{fig:sim6_example}, where we used
$L=20$ and $M=400$. Although $N=1000$, we only plotted the first $100$ signal values
$x(i)$ for ease of visualization.

To compare the performance of CS-BP with other CS decoding algorithms,
we also simulated:
({\em i}) $\ell_1$ decoding (\ref{eq:L1}) via linear programming;
({\em ii}) GPSR~\cite{GPSR2007}, an optimization method that minimizes
$\|\theta\|_1 + \mu \|y-\Phi \Psi \theta\|_2^2$;
({\em iii}) CoSaMP~\cite{Cosamp08}, a fast greedy solver;
and ({\em iv}) IHT~\cite{BlumensathDavies2008},
an iterative thresholding algorithm. We simulated all five methods
where $N=1000$, $S=0.1$, $L=20$, $\sigma_1=10$, $\sigma_0=1$, $p=525$, and the
measurements are noiseless. Throughout the experiment we ran the different methods
using the same CS-LDPC encoding matrix $\Phi$, the same signal $x$, and therefore same
measurements $y$. Figure~\ref{fig:mse_compare} plots the MMSE
decoding error as a function of $M$ for the five methods.
For small to moderate $M$, CS-BP exploits its knowledge about the approximately
sparse structure of $x$, and has a smaller decoding error.
CS-BP requires 20--30\% fewer measurements than the optimization methods
LP and GPSR to obtain the same MMSE decoding error;
the advantage over the greedy solvers IHT and CoSaMP is even
greater. However, as $M$ increases, the advantage
of CS-BP over LP and GPSR becomes less pronounced.

To compare the speed of CS-BP to other methods, we ran the same five
methods as before. In this experiment, we varied the signal length $N$
from $100$ to $10000$, where $S=0.1$, $L=20$, $\sigma_1=10$, $\sigma_0=1$,
$p=525$, and the measurements are noiseless.
We mention in passing that some of the algorithms that were evaluated
can be accelerated using linear algebra routines optimized
for sparse matrices; the improvement is quite modest, and the run-times
presented here do not reflect this optimization.
Figure~\ref{fig:timing}
plots the run-times of the five methods in seconds as a function of $N$.
It can be seen that LP scales more poorly than the other algorithms,
and so we did not simulate it for $N>3000$.\footnote{
Our LP solver is based on interior point methods.}
CoSaMP also seems to scale relatively poorly, although it is possible
that our conjugate gradient implementation can be
improved using the pseudo-inverse approach instead~\cite{Cosamp08}.
The run-times of CS-BP seem to scale somewhat better than IHT and GPSR.
Although the asymptotic computational complexity of CS-BP is good,
for signals of length $N=10000$ it is still slower than IHT and GPSR;
whereas IHT and GPSR essentially perform matrix-vector multiplications,
CS-BP is slowed by FFT computations
performed in each iteration for all nodes in the factor graph.
Additionally, whereas the choice $p=O(\sigma_1 / \sigma_0)$ yields
$O\left(\frac{\sigma_1}{\sigma_0} \log\left(\frac{\sigma_1}{\sigma_0}\right)N\log^2(N)\right)$
complexity, FFT computation with $p=525$ samples is somewhat slow.
That said, our main contribution is a computationally feasible Bayesian
approach, which allows to reduce the number of measurements
(Figure~\ref{fig:mse_compare});
a comparison between CS-BP and previous Bayesian approaches to
CS~\cite{BCS2008,BCSEx2008} would be favorable.

To demonstrate that CS-BP deals well with measurement noise,
recall the noisy measurement setting $y=\Phi x+z$
of Section~\ref{subsec:LDPC_Phi}, where
$z\sim\cal{N}$$(0,\sigma_Z^2)$ is AWGN with variance $\sigma_Z^2$. Our algorithm deals with noise
by convolving the noiseless version of the estimated pdf
(\ref{eq:constraint_to_var}) with the noise pdf.
We simulated decoding problems where $N=1000$, $S=0.1$, $L=20$,
$\sigma_1=10$, $\sigma_0=1$, $p=525$, and $\sigma_Z^2\in\{0,2,5,10\}$.
Figure~\ref{fig:noisy} plots the MMSE decoding error as a function
of $M$ and $\sigma_Z^2$. To put things in perspective, the
average measurement picks up a Gaussian term of variance
$L(1-S)\sigma_0^2=18$ from the signal. Although the decoding
error increases with $\sigma_Z^2$, as long as $\sigma_Z^2 \ll 18$ the noise has
little impact on the decoding error; CS-BP offers
a graceful degradation to measurement noise.

Our final experiment considers model mismatch where CS-BP
has an imprecise statistical characterization of the signal.
Instead of a two-state mixture Gaussian signal model
as before, where large coefficients have variance $\sigma_1^2$
and occur with probability $S$, we defined a $C$-component mixture
model. In our definition, $\sigma_0^2$ is interpreted as a
background signal level, which appears in all coefficients.
Whereas the two-state model adds a ``true signal" component of
variance $\sigma_1^2-\sigma_0^2$ to the background signal,
the $C-1$ large components
each occur with probability $S$ and the amplitudes of the
true signals are $\sigma_2,2\sigma_2,\ldots,(C-1)\sigma_2$,
where $\sigma_2$ is chosen to preserve the total signal energy.
At the same time, we did not change the signal priors in CS-BP,
and used the same two-state mixture model as before.
We simulated decoding problems where $N=1000$, $S=0.1$, $L=20$,
$\sigma_1=10$, $\sigma_0=1$, $p=525$, the measurements are noiseless,
and $C\in\{2,3,5\}$.
Figure~\ref{fig:robustness} plots the MMSE
decoding error as a function of $M$ and $C$.
The figure also shows how IHT and GPSR perform, in order to
evaluate whether they are more robust than the Bayesian approach of CS-BP.
We did not simulate CoSaMP and $\ell_1$ decoding, since their MMSE
performance is comparable to that of IHT and GPSR.
As the number of mixture components $C$ increases, the
MMSE provided by CS-BP increases. However, even for $C=3$ the sparsity rate
effectively doubles from $S$ to $2S$, and an increase in the
required number of measurements $M$ is expected.
Interestingly, the greedy IHT method also degrades significantly, perhaps
because it implicitly makes an assumption regarding the number of large
mixture components. GPSR, on the other hand, degrades more gracefully.

\section{Variations and enhancements}
\label{sec:variations}

{\bf Supporting arbitrary sparsifying basis $\Psi$}:\
Until now, we have assumed that the canonical sparsifying basis is used,
i.e., $\Psi=I$. In this case, $x$ itself is sparse. We now explain how CS-BP
can be modified to support the case where $x$ is sparse in an arbitrary
basis $\Psi$. In the encoder, we multiply the CS-LDPC matrix $\Phi$ by
$\Psi^T$ and encode $x$ as
$y=(\Phi\Psi^T)x=(\Phi\Psi^T)(\Psi \theta)=\Phi\theta$,
where $(\cdot)^T$ denotes the transpose operator.
In the decoder, we use BP to form the approximation $\widehat{\theta}$,
and then transform via $\Psi$ to $\widehat{x}=\Psi\widehat{\theta}$.
In order to construct the modified encoding matrix $\Phi\Psi^T$ and
later transform $\widehat{\theta}$ to $\widehat{x}$,
extra computation is needed; this extra cost is $O(N^2)$ in general.
Fortunately, in many practical situations $\Psi$ is structured (e.g., Fourier
or wavelet bases) and amenable to fast computation. Therefore, extending
our methods to such bases is feasible.

{\bf Exploiting statistical dependencies}:\
In many signal representations, the coefficients are not iid. For example,
wavelet representations of natural images often contain correlations between
magnitudes of parent and child coefficients~\cite{devore92im,Crouse98}.
Consequently, it is possible to decode signals from fewer measurements using
an algorithm that allocates different distributions to different
coefficients~\cite{BCDH2008,BCS2008spike}. By modifying the dependencies imposed by
the prior constraint nodes (Section~\ref{sec:subbp}), CS-BP decoding
supports different signal models.

{\bf Feedback}:\
Feedback from the decoder to the encoder can be used
in applications where measurements may be lost
because of transmissions over faulty channels. In an analogous manner to a
digital fountain~\cite{BLM02}, the marginal distributions
(\ref{eq:BP_marginal}) enable us to identify when sufficient information
for signal decoding has been received. At that stage,
the decoder notifies the encoder that decoding is complete, and
the stream of measurements is stopped.

{\bf Irregular CS-LDPC matrices}:\
In channel coding, LDPC matrices that have irregular row and
column weights come closer to the Shannon limit, because a small number of
rows or columns with large weights require only modest additional computation
yet greatly reduce the block error rate~\cite{RSU2001}. In an analogous manner,
we expect irregular CS-LDPC matrices to enable a further reduction in the number of
measurements required.

\section{Discussion}
\label{sec:discussion}

This paper has developed a sparse encoding matrix and belief propagation decoding algorithm
to accelerate CS encoding and decoding under the Bayesian framework.
Although we focus on decoding approximately sparse signals,
CS-BP can be extended to signals that are sparse in other bases,
is flexible to modifications in the signal model, and can address measurement noise.

Despite the significant benefits, CS-BP is not universal in the sense that
the encoding matrix and decoding methods must be modified in order to apply
our framework to arbitrary bases. Nonetheless, the necessary modifications
only require multiplication by the sparsifying basis $\Psi$ or its
transpose $\Psi^T$.

Our method resembles low density parity check (LDPC)
codes~\cite{Gallager62,RSU2001}, which use a sparse Bernoulli parity
check matrix. Although any linear code can be represented as a bipartite graph,
for LDPC codes the sparsity of the graph accelerates the
encoding and decoding processes. LDPC codes are celebrated for achieving rates
close to the Shannon limit. A similar comparison of the MMSE performance of CS-BP with
information theoretic bounds on CS performance is left for future research.
Additionally, although CS-BP is not guaranteed to converge,
the recent convergence proofs for LDLC codes~\cite{LDLC2008} suggest
that future work on extensions of CS-BP may also yield convergence proofs.

In comparison to previous work on Bayesian aspects of CS~\cite{BCS2008,BCSEx2008},
our method is much faster, requiring only $O(N\log^2(N))$
computation. At the same time, CS-BP offers significant flexibility, and
should not be viewed as merely another fast CS decoding algorithm.
However, CS-BP relies on the sparsity of CS-LDPC matrices, and future
research can consider the applicability of such matrices
in different applications.

\appendix

{\bf Outline of proof of Theorem~\ref{th:median_algo}}:\
The proof begins with a derivation of probabilistic bounds
on $\|x\|_2$ and $\|x\|_{\infty}$.
Next, we review a result by Wang et al.~\cite[Theorem 1]{WGR2007}.
The proof is completed by combining the bounds with the result
by Wang et al.

{\bf Upper bound on $\|x\|_2^2$}:\
Consider $\|x\|_2^2=\sum_{i=1}^N x_i^2$, where the
{\em random variable} (RV) $X_i$ has a mixture distribution
\[
X_i^2\sim \left\{
\begin{array}{ll}
\chi^2\sigma_1^2 & \mbox{w.p. }S \\
\chi^2\sigma_0^2 & \mbox{w.p. }1-S \\
\end{array}\right..
\]
Recall the {\em moment generating function} (MGF),
$M_X(t) = E[e^{tx}]$.
The MGF of a Chi-squared RV satisfies
$M_{\chi^2}(t)=(1-2t)^{-\frac{1}{2}}$.
For the mixture RV $X_i^2$,
\[
M_{X_i^2}(t) = \frac{S}{\sqrt{1-2t\sigma_1^2}}
+ \frac{1-S}{\sqrt{1-2t\sigma_0^2}}.
\]
Additionally, because the $X_i$ are iid,
$M_{\|x\|_2^2}(t)= \left[M_{X_i^2}(t)\right]^N$.
Invoking the Chernoff bound, we have
\[
\Pr\left( \|x\|_2^2 < SN\sigma_1^2 \right) <
e^{-tSN\sigma_1^2} \left[
\frac{S}{\sqrt{1-2t\sigma_1^2}}
+ \frac{1-S}{\sqrt{1-2t\sigma_0^2}}
\right]^N
\]
for $t<0$.
We aim to show that $\Pr\left(\|x\|_2^2 < SN\sigma_1^2\right)$
decays faster than $N^{-\gamma}$ as $N$ is increased. To do so, let
$t=-\frac{\alpha}{\sigma_1^2}$, where $\alpha>0$.
It suffices to prove that there exists some $\alpha$ for which
\[
f_1(\alpha) =
e^{\alpha S}
\left[
\frac{S}{\sqrt{1+2\alpha}}
+ \frac{1-S}{\sqrt{1+2\alpha\left(\frac{\sigma_0}{\sigma_1}\right)^2}}
\right] <1.
\]
Let $f_2(\alpha)= \frac{1}{\sqrt{1+2\alpha}}$
and $f_3(\alpha)= e^{\alpha}$.
It is easily seen via Taylor series that
$f_2(\alpha) = 1 - \alpha + O(\alpha^2)$ and
$f_3(\alpha) = 1+\alpha+O(\alpha^2)$, and so
\begin{eqnarray*}
f_1(\alpha) &=&
e^{\alpha S} \left[
S \left( 1 - \alpha + O(\alpha^2) \right)
+ (1-S) \left( 1
- \alpha\left(\frac{\sigma_0}{\sigma_1}\right)^2
 + O\left( \alpha^2 \left(\frac{\sigma_0}{\sigma_1}\right)^4 \right) \right)
\right]
\\
&=&
\left[ 1 + \alpha S + O(\alpha^2S^2) \right]
\left[
1 - \alpha \left(S+(1-S) \left(\frac{\sigma_0}{\sigma_1}\right)^2\right)
+O(\alpha^2) \right].
\end{eqnarray*}
Because of the negative term $-\alpha(1-S)\left(\frac{\sigma_0}{\sigma_1}\right)^2<0$,
which dominates the higher order term $O(\alpha^2)$ for small $\alpha$,
there exists $\alpha>0$, which is independent of $N$,
for which $f_1(\alpha)<1$. Using this $\alpha$, the Chernoff bound provides
an upper bound on $\Pr\left( \|x\|_2^2 < SN\sigma_1^2 \right)$
that decays exponentially with $N$. In summary,
\begin{equation}
\label{eq:bound_ell_2}
\Pr\left( \|x\|_2^2 < SN\sigma_1^2 \right) = o(N^{-\gamma}).
\end{equation}

{\bf Lower bound on $\|x\|_2^2$}:\
In a similar manner, MGF's and the Chernoff bound can be used to
offer a probabilistic bound on the number of large Gaussian
mixture components
\begin{equation}
\label{eq:bound_numlarge}
\Pr\left( \sum_{i=1}^N Q(i) > \frac{3}{2}SN  \right) = o(N^{-\gamma}).
\end{equation}
Taking into account the limited number of
large components and the expected squared $\ell_2$ norm,
$E[\|x\|_2^2]=N[S\sigma_1^2+(1-S)\sigma_0^2]$, we have
\begin{equation}
\label{eq:bound_ell_2b}
\Pr\left( \|x\|_2^2 > N[2S\sigma_1^2+(1-S)\sigma_0^2] \right) = o(N^{-\gamma}).
\end{equation}
We omit the (similar) details for brevity.

{\bf Bound on $\|x\|_{\infty}$}:\
The upper bound on $\|x\|_{\infty}$ is obtained by first considering large
mixture components and then small components.
First, we consider the large Gaussian mixture components, and
denote $x_L = \{x(i):\ Q(i)=1\}$.
\begin{eqnarray}
\Pr\left( \|x_L\|_{\infty} < \sqrt{2\ln(SN^{1+\gamma})}\sigma_1
\left| \sum_{i=1}^N Q(i) \leq \frac{3}{2}SN \right.\right)
&\geq&
\left[ f_4\left(\sqrt{2\ln(SN^{1+\gamma})}\right) \right]^{\frac{3}{2}SN}
\label{eqn:bound2_ineq1} \\
&>&
\left[ 1-\frac{ f_5\left(\sqrt{2\ln(SN^{1+\gamma})}\right) }
{\sqrt{2\ln(SN^{1+\gamma})}} \right]^{\frac{3}{2}SN}
\label{eqn:bound2_ineq2} \\
&>&
1 - \frac{3}{2}SN
\frac{ f_5\left(\sqrt{2\ln(SN^{1+\gamma})}\right) }
{\sqrt{2\ln(SN^{1+\gamma})}}
\label{eqn:bound2_ineq3} \\
&=&
1 - \frac{3SN} {2\sqrt{2\ln(SN^{1+\gamma})}}
\frac{e^{-\frac{1}{2} 2\ln(SN^{1+\gamma}) }}{\sqrt{2\pi}}
\nonumber \\
&=&
1 - \frac{3N^{-\gamma}} {4\sqrt{\ln(SN^{1+\gamma})}},
\nonumber
\end{eqnarray}
where
$f_4(\alpha)=\frac{1}{\sqrt{2\pi}} \int_{-\infty}^{\alpha} e^{-u^2/2} du$
is the cumulative distribution function of the standard normal distribution,
the inequality (\ref{eqn:bound2_ineq1})
relies on $f_4(\cdot)<1$ and the possibility that $\sum_{i=1}^N Q(i)$
is strictly smaller than $\frac{3}{2}SN$,
$f_5(\alpha)=\frac{1}{\sqrt{2\pi}}e^{-\alpha^2/2}$ is the
pdf of the standard normal distribution,
(\ref{eqn:bound2_ineq2}) relies on the bound
$f_4(\alpha)>1-f_5(\alpha)/\alpha$,
and the inequality (\ref{eqn:bound2_ineq3}) is motivated by
$(1-\alpha)^{\beta}>1-\alpha\beta$ for $\alpha,\beta>0$.
Noting that $\ln(SN^{1+\gamma})$ increases with $N$,
for large $N$ we have
\begin{equation}
\Pr\left( \|x_L\|_{\infty} < \sqrt{2\ln(SN^{1+\gamma})}\sigma_1
\left| \sum_{i=1}^N Q(i) \leq \frac{3}{2}SN \right.\right)
> 1 - \frac{N^{-\gamma}} {5}.
\label{eq:ell_infty_xL}
\end{equation}
Now consider the small Gaussian mixture components, and
denote $x_S = \{x(i):\ Q(i)=0\}$. As before,
\begin{eqnarray}
\Pr\left( \|x_S\|_{\infty} < \sqrt{2\ln(SN^{1+\gamma})}\sigma_1 \right)
&\geq&
\left[ f_4\left(\sqrt{2\ln(SN^{1+\gamma})} \frac{\sigma_1}{\sigma_0} \right) \right]^N
\label{eqn:bound3_ineq1} \\
&>&
1 - \frac{N} {\sqrt{2\ln(SN^{1+\gamma})} \frac{\sigma_1}{\sigma_0} }
\frac{e^{-\frac{1}{2} 2\ln(SN^{1+\gamma}) \left(\frac{\sigma_1}{\sigma_0}\right)^2 }}{\sqrt{2\pi}},
\nonumber
\end{eqnarray}
where in (\ref{eqn:bound3_ineq1}) the number of small mixture
components is often less than $N$. Because $\sigma_1>\sigma_0$,
for large $N$ we have
\begin{equation}
\Pr\left( \|x_S\|_{\infty} < \sqrt{2\ln(SN^{1+\gamma})}\sigma_1 \right)
> 1 - \frac{N^{-\gamma}} {5}.
\label{eq:ell_infty_xS}
\end{equation}
Combining (\ref{eq:bound_numlarge}), (\ref{eq:ell_infty_xL}) and
(\ref{eq:ell_infty_xS}), for large $N$ we have
\begin{equation}
\label{eq:bound_ell_infty}
\Pr\left( \|x\|_{\infty} < \sqrt{2\ln(SN^{1+\gamma})}\sigma_1 \right)
> 1 - \frac{N^{-\gamma}} {2}.
\end{equation}

{\bf Result by Wang et al.~\cite[Theorem 1]{WGR2007}}:\

\begin{THEO}[\cite{WGR2007}] \label{th:Wang}
Consider $x\in\mathbb{R}^N$ that satisfies
the condition
\begin{equation}
\label{eq:def_Q}
\frac{\|x\|_{\infty}}{\|x\|_2} \leq Q.
\end{equation}
In addition, let $V$ be any set of $N$ vectors
$\{v_1,\ldots,v_N\}\subset \mathbb{R}^N$.
Suppose a sparse random matrix $\Phi\in\mathbb{R}^{M \times N}$
satisfies
\[
E[\Phi_{ij}]=0,
E[\Phi^2_{ij}]=1,
E[\Phi^4_{ij}]=s,
\]
where $\frac{1}{s}=\frac{L}{N}$ is the fraction of non-zero entries
in $\Phi$. Let
\begin{equation}
\label{eq:th_Wang_M}
M = \left\{
\begin{array}{ll}
O\left( \frac{1+\gamma}{\epsilon^2}sQ^2\log(N) \right) & \mbox{if $sQ^2\geq\Omega(1)$} \\
O\left( \frac{1+\gamma}{\epsilon^2}\log(N) \right) & \mbox{if $sQ^2\leq O(1)$} \\
\end{array}
\right..
\end{equation}
Then with probability at least $1-N^{-\gamma}$, the random projections
$\frac{1}{M}\Phi x$ and $\frac{1}{M}\Phi v_i$ can produce an
estimate $\widehat{a}_i$ for $x^Tv_i$ satisfying
\[
|\widehat{a}_i-x^Tv_i| \leq \epsilon \|x\|_2\|v_i\|_2,\
\forall i\in\{1,\ldots,N\}.
\]
\end{THEO}

{\bf Application of Theorem~\ref{th:Wang} to proof of Theorem~\ref{th:median_algo}}:\
Combining (\ref{eq:bound_ell_2}), (\ref{eq:bound_ell_2b}),
and (\ref{eq:bound_ell_infty}), the union bound demonstrates that
with probability lower bounded by $1 - N^{-\gamma}$
we have $\|x\|_{\infty} < \sqrt{2\ln(SN^{1+\gamma})} \sigma_1$
and $\|x\|_2^2 \in (NS\sigma_1^2,N[2S\sigma_1^2+(1-S)\sigma_0^2])$.\footnote{
The $o(\cdot)$ terms (\ref{eq:bound_ell_2})
and (\ref{eq:bound_ell_2b}) demonstrate that there exists some $N_0$ such that
for all $N>N_0$ the upper and lower bounds on $\|x\|_2^2$ each hold with
probability lower bounded by $1-\frac{1}{4}N^{\gamma}$, resulting
in a probability lower bounded by $1 - N^{-\gamma}$ via the union bound.
Because the expression (\ref{eq:th:median_algo}) for the number of
measurements $M$ is an order term, the case where $N\leq N_0$ is inconsequential.}
When these $\ell_2$ and $\ell_{\infty}$ bounds hold, we can apply Theorem~\ref{th:Wang}.

To apply Theorem~\ref{th:Wang}, we must specify
({\em i}) $Q$ (\ref{eq:def_Q});
({\em ii}) the test vectors $(v_i)_{i=1}^N$;
({\em iii}) the matrix sparsity $s$; and
({\em iv}) the $\epsilon$ parameter.
First, the bounds on $\|x\|_2$ and $\|x\|_{\infty}$ indicate that
$\frac{\|x\|_{\infty}}{\|x\|_2} \leq Q =
\sqrt{ \frac{ 2\ln(SN^{1+\gamma})}{SN}}$.
Second, we choose $(v_i)_{i=1}^N$ to be the $N$ canonical vectors of
the identity matrix $I_N$, providing $x^Tv_i=x_i$.
Third, our choice of $L$ offers
$s=\frac{N}{L}=\frac{NS}{\eta\ln(SN^{1+\gamma})}$.
Fourth, we set
\[
\epsilon = \frac{\mu\sigma_1}
{\sqrt{N[2S\sigma_1^2+(1-S)\sigma_0^2]}}.
\]
Using these parameters, Theorem~\ref{th:Wang} demonstrates that all $N$
approximations $\widehat{a}_i$ satisfy
\[
|\widehat{a}_i-x_i|
= |\widehat{a}_i-x^Tv_i|
\leq \epsilon \|x\|_2\|v_i\|_2
< \mu\sigma_1
\]
with probability lower bounded by $1-N^{-\gamma}$.
Combining the probability that the $\ell_2$ and $\ell_{\infty}$
bounds hold and the decoding probability offered by
Theorem~\ref{th:Wang}, we have
\begin{equation}
\label{eq:ell_infty_quality}
\|\widehat{a}-x\|_{\infty} < \mu \sigma_1
\end{equation}
with probability lower bounded by $1-2N^{-\gamma}$.

We complete the proof by computing
the number of measurements $M$ required (\ref{eq:th_Wang_M}). Because
$sQ^2= \frac{K}{\eta\ln(SN^{1+\gamma})}
\frac{ 2\ln(SN^{1+\gamma})}{SN} = \frac{2}{\eta}$, we need
\[
M=O\left( (1+2\eta^{-1})\frac{1+\gamma}{\epsilon^2}\log(N) \right)
=
O\left(
N(1+2\eta^{-1})\frac{(1+\gamma)}{\mu^2}\left[2S+(1-S)\left(\frac{\sigma_0}{\sigma_1}\right)^2\right]
\log(N)
\right)
\]
measurements.
\hfill$\Box$

\section*{Acknowledgments}
Thanks to David Scott, Danny Sorensen, Yin Zhang, Marco Duarte,
Michael Wakin, Mark Davenport, Jason Laska, Matthew Moravec, Elaine Hale,
Christine Kelley, and Ingmar Land for informative and inspiring conversations.
Thanks to Phil Schniter for bringing his related work~\cite{FBMP2009} to
our attention.
Special thanks to Ramesh Neelamani, Alexandre de Baynast, and
Predrag Radosavljevic for providing helpful suggestions for implementing BP;
to Danny Bickson and Harel Avissar for improving our implementation;
and to Marco Duarte for wizardry with the figures.
Additionally, the first author thanks the Department of Electrical Engineering
at the Technion for generous hospitality while parts of the work were being
performed, and in particular the support of Yitzhak Birk and Tsachy Weissman.
Final thanks to the anonymous reviewers, whose superb comments helped
to greatly improve the quality of the paper.

\bibliography{csbp062009arxiv.bib}

\end{document}